\documentclass{article}
\usepackage[english]{babel}

\usepackage{epsfig}

\usepackage{amsfonts,amssymb,amsmath}

\usepackage[utf8]{inputenc}
\usepackage[margin=1.4in]{geometry}

\usepackage[T1]{fontenc}

\graphicspath{{figs/}} 

\usepackage[font={sl}, labelfont={bf,sc}, margin=1.3cm]{caption}

\begin{document}

\title{Are Triggering Rates of Labquakes Universal? Inferring Triggering Rates From Incomplete Information.}
\author{
Jordi Bar\'o, J\"orn Davidsen\\
\small\textit{Complexity Science Group, Department of Physics and Astronomy, University of Calgary, Canada}
}

\maketitle

\abstract{ 
The acoustic emission activity associated with recent rock fracture experiments under different conditions has indicated that some features of event-event triggering are independent of the details of the experiment and the materials used and are often even indistinguishable from tectonic earthquakes. While the event-event triggering rates or aftershock rates behave pretty much identical for all rock fracture experiments at short times, this is not the case for later times. Here, we discuss how these differences can be a consequence of the aftershock identification method used and show that the true aftershock rates might have two distinct regimes. Specifically, tests on a modified Epidemic Type Aftershock Sequence model show that the model rates cannot be correctly inferred at late times based on temporal information only if the activity rates or the branching ratio are high. We also discuss both the effect of the two distinct regimes in the aftershock rates and the effect of the background rate on the inter-event time distribution. Our findings should be applicable for inferring event-event triggering rates for many other types of triggering and branching processes as well.
}

\section{Event-event triggering \& aftershocks}

Many striking features of physical, geophysical, biological or social processes can be portrayed as patterns or clusters of localized events. Specific examples include magnetization processes~\cite{sethna01,durin06,papanikolaou11}, martensitic transformations~\cite{baro2014}, fracture processes~\cite{bonamy11,tantot13,baro2013,kun14,maekinen15,ribeiro15}, natural or induced earthquakes~\cite{benzion08,gu13,davidsen16m,maghsoudi16}, solar flares~\cite{baiesi06m,arcangelis06a}, extreme bursts in the solar wind~\cite{moloney11,moloney13}, the spread of infections~\cite{turcotte99}, extinctions of species~\cite{bak,crutchfield,drossel01}, neural spikes~\cite{friedman12,yaghoubi17}, booms and bursts of markets and economies~\cite{bak,sornette1996,farmer05,lillo03,weber07,petersen10,siokis12}, media coverage~\cite{klimek11} -- to name a few. A generic attribute in all these cases is that one event can trigger or somehow induce another one to occur -- or possibly numerous further events. One of the most prominent examples of such event-event triggering are \textit{aftershocks}~\cite{weiss04,crane08,sornette09,gu13amj,hainzl14,stojanova14,davidsen16m}. Aftershock sequences are characterized by time-varying (local) event rates, which are often empirically found to approximately follow --- across a wide range of scales and systems from friction and fracture to socio-economic systems~\cite{gu13amj,utsu95,moradpour14,davidsen14,goebel13a,baro2013,maekinen15,ribeiro15,arcangelis06a,lillo03,weber07,petersen10,siokis12,klimek11} --- the Omori-Utsu (OU) relation,
\begin{equation}
r(t) = \frac{K}{(t + c)^p} \equiv \frac{1}{\tau (t / c + 1)^p },
\label{eq:BD:ou}
\end{equation}
first proposed for earthquakes~\cite{omori}. Here, $t$ measures the time after the triggering event, $p$ is typically close to 1 ($p\gtrsim 1$ if one only considers directly triggered events~\cite{davidsen14}) and $\tau \equiv {c^p}/{K}$. $K$ is typically found to increase with the energy of the trigger though the exact dependence of $K$, $c$ and hence $\tau$ on different parameters is an active field of research~\cite{davidsen16m}.\\

The OU relation with $p \approx 0.7$ has in particular been observed in acoustic emission (AE) experiments of rock fracture across a range of different materials and conditions~\cite{baro2013,castillo2013,nataf2014}, denoting some sort of universality in the response of disordered materials under mechanical stress, an hypothesis already suggested by the scale invariance in other physical and statistical relations~\cite{davidsen05sg,davidsen2013,goodfellow2014}. Yet, a $p$-value significantly lower than 1 implies that the number of events directly or indirectly triggered by a single event is infinite. Indeed, a more recent study of rock fracture experiments using a more reliable technique to identify triggered events has shown that there are significant deviations from the OU relation at late times with a steeper decay that ensures that the number of triggered events is finite~\cite{davidsen17}. Identifying the aftershocks is a general challenge for all triggering processes since a detailed or ``fundamental'' knowledge of the underlying microscopic dynamics and causal information is typically not available~\cite{davidsen06pm}. In the case of earthquakes, the most reliable methods to identify triggering relations use spatio-temporal correlations between events~\cite{gu13amj,moradpour14,marsan08,zaliapin13a}. In the absence of spatial information --- as it is the case for previous rock fracture experiments~\cite{baro2013,castillo2013,nataf2014} --- this is not an option and one has to rely on  the measurement of the whole activity rate after each event~\cite{baro2013,helmstetter2003}. This technique can lead to a strong bias in the estimation of triggering rates in cases where either the number of triggered events or the background rate of events activated by other mechanisms is high, or both, as we show explicitly here.\\

Specifically, in this paper we aim to quantify the bias of this technique and establish under which conditions it can serve as a reasonable estimator for the triggering rates. We test it against synthetic catalogs generated by a modified Epidemic-Type Aftershock Sequence (ETAS) model with a triggering rate characterized by two power laws, as observed in the most recent rock fracture experiments~\cite{davidsen17}. 
We address the possibility that the low experimental OU exponent ($p$ values) and some other inconsistent results observed in the previous rock fracture experiments~\cite{baro2013,castillo2013,nataf2014} are a consequence of the hidden complexity of the triggering process. 
First, we formulate triggering in terms of branching processes. 
The true \textit{direct} and \textit{compound triggering rates} can be estimated reliably only under certain conditions. Instead, one is often limited to less reliable estimators such as those based on the \textit{mean aftershock sequence rates} (MASR) and the \textit{distribution of waiting times} (DWT) to extract the properties of the triggering process. We discuss the limitations of such methods.
Next, we introduce the modified ETAS model with the triggering rates characterized by two power laws as observed in rock fracture experiments with spatio-temporal information \cite{davidsen17}. Considering that only magnitude and temporal information is available, we interpret the measured MASR in terms of direct and compound triggering rates, and identify the parameters giving rise to different power-law regimes in the DWT. Finally, we discuss whether these numerical results can provide an explanation for anomalous features in the experimental results.

\section{Event-event triggering represented as a branching process}

The study of systems exhibiting localized events and triggering between them can be cast in the language of point processes~\cite{kagan87,ogata1999,daley2007,corral2009,davidsen16m}. A stochastic point process is fully determined by a function called the intensity, which quantifies the probability of occurrence of an event of size $M$ at time $t$ and at location $\vec{z}$:\\
\begin{equation}
\mu(t,\vec{z},M) := \mathrm{Prob}\left\lbrace{\mathrm{event }\:\mathrm{of }\:\mathrm{size }\:M\: \mathrm{at }\: t,\vec{z}}\right\rbrace \:dt \: d\vec{r}\: dM.
\end{equation}

The measured activity rate can vary over time due to an explicit temporal variation of external parameters, and/or as a consequence of previous activity. In the latter case, the intensity depends explicitly on the history of the point process ($\mathcal{H}_t:=\lbrace \mathrm{all}\:\mathrm{events}\: i \: ;  t_i < t \rbrace$).  The exact intensity ($\mu(t,\mathcal{H}_t)$) of a process exhibiting triggering involves in general a  complex contribution of the whole history $\mathcal{H}_t$.
In a simplified approach, the contribution to the intensity of each past event can be linearized in a Hawkes self-exciting point process~\cite{hawkes1974}: 
\begin{equation}
    \mu (t,\mathbf{z},m | \mathcal{H}_t) = \mu_0(t,m) + \sum_{i \in \mathcal{H}_t}{\phi(t-t_i,\vec{z}-\vec{z}_i,m|M_i)}
    \label{eq:BD:hawkes}
\end{equation}
Stochastic Hawkes processes can be reinterpreted as the outcome of branching processes where each event is either a background event or has a single parent. Given a background event (G=0), a sequence of first generation events (G=I) can be triggered, after a time difference $\tau$ and at relative position $\vec{r}$ from the background event, according to an intensity factor represented by the \textit{triggering kernel} $\phi(\tau,\vec{r},m|M_i)$. Each event in the first generation can itself trigger a sequence of second generation events (G=II) with the same relative kernel $\phi(\tau,\vec{r},m|M_i)$, and so on until a whole triggering cascade or tree is generated up to some $n$-th generation that does not trigger further events. 
The major physical constraint to the model is the stability of the branching process, requiring that the average number of events directly triggered from a single parent ---computed as the \textit{average branching ratio}: $n_b= \left\langle{ \int d\tau \int d\vec{r} \int d{m} \int dm' \; \phi(\tau,\vec{r},m'|m) }\right\rangle $--- has to be lower than one.
Triggering trees can be spatially and temporally overlapping, generating a complex triggering forest difficult to disentangle in practice.\\

\subsection{Direct and compound event-event triggering rates}

The linear Hawkes model is useful for both the development of forecasting tools \cite{helmstetter2006,filimonov2012,tiampo12,bacry2015} and the deep understanding of the fundamental physics behind avalanche processes~\cite{sibani2005,van2013,Jagla2014,janicevic2016}. 
Both purposes require a reliable estimation of the triggering kernel. 
If the actual pairwise parent-child relations are retrievable, or can be estimated from declustering techniques~\cite{zaliapin08,marsan08,gu13amj,zaliapin13a,hainzl2016}, one can measure the \textit{direct} (or \textit{bare}~\cite{Helmstetter2002}) triggering rates and use them as a good estimator for the triggering kernel: $\widehat{\phi}(\tau,\vec{r},m'|m)$.
\\

In some cases one has to deal with time series without spatial information, or situations where the spatial kernel is too spread to retain meaningful information. One is left with the marginal temporal point process with a triggering kernel $\phi(\tau,m | m_0) $. The \textit{compound} (or \textit{dressed}~\cite{Helmstetter2003b}) triggering rates are the expected temporal activity during the span of a triggering cascade generated from a background event (the root of the tree) of magnitude $m_0$  at $t_0$. The compound triggering rates can be computed from the direct triggering rates as:
\begin{equation}
    \Phi_{c}(\tau,m | m_0) := \sum_{G=I, II, ...}^{\infty} {\Phi_{G}(\tau,m | m_0)}
\label{eq:BD:M2}
\end{equation}
where $\tau=t-t_0$ and $\Phi_{G=I}$ corresponds to the direct rates from the background event, $\Phi_{G=II}$ designates the rate of events triggered by all first generation events, etc..
Under certain conditions one can calculate the compound rates analytically, or at least find an approximate solution, as for the case exposed below.
The magnitude of triggered events is usually assumed as an independent variable: $\phi(\tau,m|m_0)= \rho(m)\phi(\tau|m_0)$, where $\rho(m)$ is the distribution of magnitudes.
In such cases, the intensity of the first generation ($G=I$), equivalent to the kernel from the triggering background event, can be written as:
\begin{equation}\Phi_{I}(\tau,m|m_0) \equiv  \rho(m)\phi(\tau|m_0)
\label{eq:BD:G1}
\end{equation}
and  each one of the higher order generations contribute to the intensity as:
\begin{equation}
    \Phi_{G}(\tau,m|m_0) =  \rho(m) \int_{m_{c}}^{\infty} dm'  \int_{0}^{\tau} dt' \Phi_{G-1}(t'|m_0) \phi(\tau-t'|m') \rho(m') 
\label{eq:BD:GN}
\end{equation}
where $t'$ and $m'$ are the occurrence time and magnitude of the events from the $(G-1)$'th generation originated by the mainshock ($m_{0},t_{0}$), and $m'$ is distributed according to $\rho(m')$. 
If the kernel can be further separated between the temporal  and productivity term: $\phi(\tau|m'):= k(m')\widetilde{\phi}(\tau)$, the branching ratio is simplified as: $n_b= \int  k(m')\;\rho(m')\; dm'$, and each element is directly triggered by the previous generation:
\begin{equation}
    \Phi_{G}(\tau|m_0)  
    = n_b \int_{0}^{\tau} dt'  \Phi_{G-1}(t',m_0) \widetilde{\phi}(\tau-t') 
\end{equation}
Substituting Eq.~\eqref{eq:BD:G1} in the generic recurrence the triggering rates of generation $G$ can be expressed as a series of convolution operations (*) as: 
\begin{equation}
\begin{array}{rl}
    \Phi_{G}(\tau|m_0) & = 
    K(m_{0})
    {n_{b}}^{g} \left({\widetilde{\phi}_{1} * \widetilde{\phi}_{2} *... * \widetilde{\phi}_{g}}\right)
\end{array} 
\end{equation}
This recurrent expression can be solved in the Laplace transformed space. Given $\widetilde{\psi}(s) :=
\mathcal{L}\left({\widetilde{\phi}(y)}\right)$
 the compound rates can be found as: 
\begin{equation}
    \Phi_{c}(\tau|m_0) =    {K}(m_0) \sum_{g=1}^{\infty} \mathcal{L}^{-1} \left({
    \left({n_{b} 
    \widetilde{\psi}(s)}\right)^{g}
    }\right)
    \label{eq:BD:M2solution}
\end{equation}

\subsection{Mean aftershock sequence rates}
In the absence of any information regarding the precise topological structure of the branching process we can still try to estimate the triggering kernel under certain conditions.
The branching nature of the model, and the independence between terms, impose Markovian correlations in the parent-child relationship. All triggering branches are independent and, thus, triggering trees can be considered independently of the generation of the parent without loss of generality. Under these premises, we can consider all triggering trees originated from any event as statistically equivalent. 
Thus, we can measure  the activity rates conditioned to the presence of a trigger or mainshock ($t_0,m_0$) --- the \textit{mean aftershock sequences rates} (MASR) --- as the expected density of events: 
\begin{equation}
    \mathrm{MASR}(\tau,m|m_0) := \left\langle{\rho(t-t_i=\tau,m,M_i=m_0)}\right\rangle  
\end{equation}
In general, the measurement of MASR corresponds to the compound rates plus the contribution of all events without a causal connection to the mainshock. If the branching ratio is low, we can consider this second contribution as the independent activity:
\begin{equation}
    \mathrm{MASR}(\tau|m_0) \approx  \Phi_{c}(\tau|m_0)+
        \left\langle{ \mu_0(t)+ 
    \int dt' 
    \mu(t') \phi(t-t'|m_0) }\right\rangle
    \approx \Phi_{c}(\tau|m_0) +  \langle \mu(t) \rangle  
    \label{eq:BD:M3}
\end{equation}
Under certain conditions the MASR serve as a good approximation to the compound or even the direct rates. If the background activity is low enough to isolate independent triggering trees, the contribution of independent events is small and $\mathrm{MASR}(\tau|m_0) \approx  \Phi_{c}(\tau|m_0)$. Furthermore, if the branching ratio is low enough to neglect secondary triggering as a major contribution to the triggering rates, compound and direct rates are similar and hence $\Phi_{c}(\tau|m_0) \approx  \phi(\tau|m_0)$.

\subsection{The Distribution of Waiting Times}

Finally, we shall mention the more naive approach to study triggering based on the \textit{distribution of waiting times} (DWT) or times between consecutive events. 
Since the DWT is a memoryless measurement, different point process (with and without correlations) can give rise to similar distributions. It is advised to use the DWT with caution when assessing the presence of triggering, and to use more reliable techniques instead~\cite{van2010,bi1989,carbone2006,baro2016,baro2016b}.\\

Specifically, different phenomena can give rise to power-law regimes in the DWT.
For a Poisson process, all events are independent and their DWT renders a decaying exponential with a characteristic rate (e.g. $\mu_{0}$ in Eq.~\ref{eq:BD:hawkes}). Yet, if this rate varies over time, the compound DWT corresponds to a superposition of exponential distributions.
If the rate starts from zero ($\mu_{0}(t=0)=0$) and increases with time, we can always expand the temporal dependence of the background rate in a power series around the origin: $\lim_{t\to 0^{+}}\mu_{0}(t) \approx t^{1/\xi}$. It can be shown that, for long waiting times, the DWT for this process of independent events will be: $ \mathrm{DWT}(\delta) d\delta \propto \delta^{-\xi-2} d\delta $~\cite{baro2013,wheatland2000,corral2003,shcherbakov2005}.\\

On the other hand, a power-law DWT can also emerge from triggering. Typically, the presence of triggering is identified as an anomalous behavior in the distribution of short waiting times. Direct triggering rates decaying in time as a power-law $r(t-t_i) \propto (t-t_i)^{-p}$ ---such as the Omori-Utsu relation in Eq.~\eqref{eq:BD:ou}--- return also a power-law with an exponent $2-1/p$~\cite{corral2003,shcherbakov2005,utsu95}. However, some experimental measurements also display discrepancies with this exact relation \cite{baro2013,baro2014,makinen2015}. Finally, when power-law triggering processes coexist with a time-dependent background rate we can find a double power-law DWT fulfilling certain scaling relations~\cite{corral2003,baro2013}.

\section{Triggering Models from Empirical Data}
One of the most studied triggering processes are aftershock sequences in seismology. Seismic activity increases after major earthquakes, a phenomenon consistent with the idea that this event --- often called a main shock --- triggered other ones. 
In 1894 F. Omori realized that the rate of the triggered earthquakes ---the aftershocks--- after the 1891 Nobi main shock decayed in time following a power-law relation with exponent $p\approx 1$ \cite{omori}. The modified Omori-Utsu relation~\cite{utsu95}, as stated in Eq.~\eqref{eq:BD:ou}, presents a good approximation for the activity rates after almost all major earthquakes in recent history.
The size of an earthquake is quantified by the magnitude $m$, a logarithmic measure of the seismic moment released by the slip associated with an earthquake. In 1944 Gutenberg and Richter \cite{gutenberg1944} established that the number of earthquakes above a certain magnitude $m$ approximately behaves as $N_{>}(m) \sim 10^{-b m}$, equivalent to a power-law distribution of the seismic moment or an exponential distribution of magnitudes:
\begin{equation}
\rho(m)dm = 
    {b\ln (10)} 10^{-b (m-m_{c})} dm \; \quad \mathrm{for} \quad m_{c}\leq m
\label{eq:BD:gr}
\end{equation}
where $m_{c}$ corresponds to the magnitude of completeness of the given catalog. The value of $b$ is close to unity and Eq.~\eqref{eq:BD:gr} extends down to magnitudes as low as $m=-4.4$~\cite{kwiatek10}. Typically, magnitudes can be considered to be independent~\cite{davidsen11,davidsen12}.
There exists, however, a well established relationship between the magnitude of main shocks and the number of their aftershocks $N_{\mathrm{AS}}$. The productivity relation of aftershocks states that $N_{\mathrm{AS}}$ scales with the magnitude of the mainshock $m_{0}$ as \cite{utsu95}:
\begin{equation}
N_{\mathrm{AS}}(m_{0})\propto10^{\widetilde{\alpha}}m_{0}
\label{eq:BD:productivity}
\end{equation}
This relation implies that the parameter $K$ in the OU relation~\eqref{eq:BD:ou} is not a constant and instead scales with the magnitude of the mainshock as: 
$  K(m_{0}) = k 10^{\alpha m_{0}}
  \label{eq:BD:mKernel}
$ and, in this case, $\alpha\equiv \widetilde{\alpha}$.
Yet, recent studies of earthquake catalogs indicate that the productivity relation and the Omori-Utsu relation might need to be augmented and $\alpha \neq \widetilde{\alpha}$~\cite{davidsen16m}.

\subsection{The Epidemic-Type Aftershock Sequence (ETAS) model of earthquakes}

The three statistical relations of seismology stated in Eqs.~(\ref{eq:BD:ou},\ref{eq:BD:gr},\ref{eq:BD:productivity}) can be used to define a branching process
\footnote{Notice that the exponent $1+\theta$ and the $p$ in the OU relation~\eqref{eq:BD:ou} measured from the compound rates may differ~\cite{helmstetter2003}, as can be derived from Eq.~\eqref{eq:BD:M2solution}.}, commonly known as the Epidemic Type Aftershock (ETAS) Model \cite{ogata1988}, where all the explicit dependences can be separated as:
\begin{equation}
    \phi(m,\tau,\mathbf{r}|m_{0})  = \rho(m)K(m_{0})\widetilde{\phi}_{T}(\tau)\widetilde{\phi}_{R}(\mathbf{r})
\;\;\mathrm{with}\;
\left\lbrace{
\begin{array}{lcl}
    \rho(m) &  = & b \ln (10) 10^{b(m_{c}-m)}\\
   K(m_{0}) & =  &   k 10^{\alpha m_{0}} \\
    \widetilde{\phi}_{T}(\tau) & = & \theta{C^{\theta}}{(C+\tau)^{-1-\theta}}
\end{array}
}\right.
\label{eq:BD:stdETAS}
\end{equation}

where $m_{0}$ is the magnitude of the mainshock and $\tau=t-t_{0}$. The average branching ratio is given by: $n_{b}=k\frac{b}{b-\alpha}10^{\alpha m_{c}}$. In the branching process approach, the number of triggered events is sampled as a Poisson variable with rate $K(m_{0})$.

\subsection{The modified ETAS model for rock fracture}

Many physical processes exhibit statistical features similar to those summarized in Eqs.~(\ref{eq:BD:ou}, \ref{eq:BD:gr}, \ref{eq:BD:productivity}) for seismicity. Specifically, the ETAS model describes remarkably well some aspects of the temporal sequences of acoustic emission (AE) events recorded during the failure of rocks and porous materials under compression \cite{baro2013}.
However, the full spatio-temporal triggering cascades, which have only become accessible very recently, in AE experiments of rock fracture reveal a more complex triggering kernel than the standard ETAS model~\cite{davidsen17}. 
While the event magnitudes appear to be independent ($\phi(m,\tau|m_{0})  = \rho(m)\phi(\tau|m_{0})$), empirical evidence suggests that there is a characteristic time associated with the triggering rates that scales with the magnitude of the mainshock. Similar behavior has also been observed very recently in earthquake catalogs from Southern California~\cite{davidsen14,davidsen16m}. Thus, the term $\phi(\tau|m_{0})$ cannot be separated in independent terms $K(m_{0})\phi(\tau)$. Specifically, the following scaling form has been observed: 
$\phi(\tau| m_0)  \sim 10^{\alpha m_0}  \widetilde{\phi}(10^{-\alpha_{\tau} m_0}\tau)$ with $\alpha_{\tau}=0.5$~\cite{davidsen17}. 
Furthermore, the short time regime, which is constant in the standard ETAS model, is better fitted by a generic power law with an exponent $0\leq p_1 < 1$. In summary, the scaling function exhibits a transition from this power-law regime $\widetilde{\phi}(x)\sim x^{-p_{1}}$ below a characteristic value $x_{c}$ of rescaled time, towards the standard $\widetilde{\phi}(x)\sim x^{-p_{2}}$ with $p_{2}>1$. 
In order to implement a modified ETAS model able to reproduce the empirical observations, we need to impose a branching ratio $n_{b}\leq 1$ for stability reasons, and set $p_{1}<1$ and $p_{2}>1$ to be integrable over the whole temporal domain. We define the positive parameters $\theta_{1}:=1-p_{1}$ and $\theta_{2}:=p_{2}-1$ for convenience.
By normalizing the kernel and imposing the continuity constrain at $x_{\pm}=x_{c}$ the scaling relation has the explicit form:
\begin{equation}
\frac{1}{x_{c}}    
\widetilde{\phi}(x/x_{c})d x = \frac{dx}{x_{c}} \left({\frac{\theta_1 \theta_2}{\theta_2 + \theta_1}}\right)
\left\lbrace{
\begin{array}{ll}
\left({ \frac{x_{c}}{x}}\right)^{1-\theta_1} \quad & \mathrm{for}\quad 0 < x/x_c\leq 1\\
\\
\left({ \frac{x_{c}}{x}}\right)^{1+\theta_2}  \quad & \mathrm{for}\quad x/x_c\geq 1\\
\end{array}
}\right.
\label{eq:BD:Kernel}
\end{equation}
Considering the same $K(m_{0})$ used in Eq.~(\ref{eq:BD:stdETAS}) and the definition of the productivity law stated in Eq.~\eqref{eq:BD:productivity}, the total number of events triggered by a mainshock of magnitude $m_0$ now scales with the compound productivity exponent $\widetilde{\alpha}=\alpha+\alpha_{\tau}$. Given the temporal and mainshock-magnitude dependent kernel ${\phi}(\tau|m_{0})$ the scaling function $\widetilde{\phi}(x/x_{c})$ from Eq.~\eqref{eq:BD:Kernel} can be retrieved as:
\begin{equation}
    \widetilde{\phi}\left({\frac{10^{-\alpha_{\tau} m_{0}}\tau}{x_{c}}}\right)  = \frac{x_{c}}{k} 10^{-\alpha m_{0}} {\phi}(\tau|m_{0}).
    \label{eq:BD:scaling}
\end{equation}
This is the form we focus on in the remainder of the paper, especially in the figures. Given the intensity of the process $\mu(t, \mathcal{H}_{t})$ from Eq.~\eqref{eq:BD:hawkes}, the terms $k$, $x$, $x_{c}$ and $\mu_{0}$ involve a temporal scale. Here, we select $\langle \mu_{0}(t) \rangle$ as our time unit. Thus, we express the parameter $x_{c}$ in units of the mean background rate: $[x_{c}] = \langle \mu_{0}(t) \rangle ^{-1}$.\\

\begin{figure}
\centering
\includegraphics{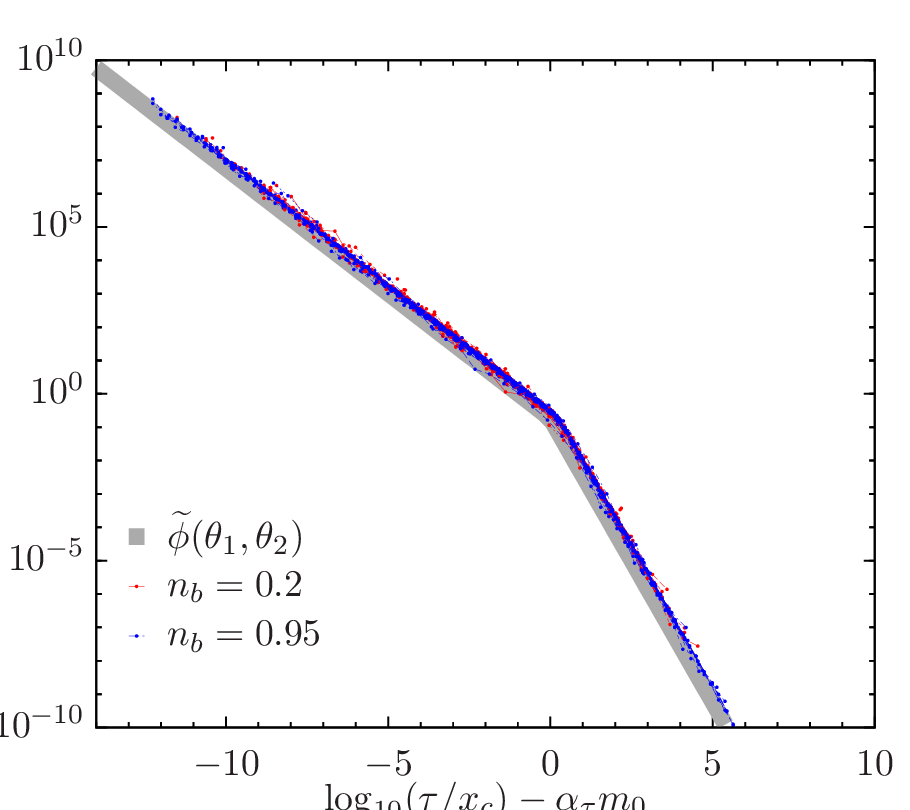}
\caption{The triggering kernel recovered by the bare triggering rates measured in the simulations, exhibiting the scaling relation in $m_0$, $x_c$ and $n_b$ summarized by Eq.~\eqref{eq:BD:scaling}. Each curve (around 7 for each simulation) corresponds to the average 1st generation activity after mainshocks in different magnitude ranges $m < M_i < m + \Delta m$, with $\Delta m = 0.5$. Six simulations are shown with values $x_c=1$, $0.0166$ and $10^{-8}$, respectively, and $n_b=0.2$ (in red) and $0.95$ (in blue), respectively.}
\label{fig:BD:M1}       
\end{figure}

We implement the ETAS model with the modified temporal kernel as defined in Eq.~\eqref{eq:BD:Kernel} and performed simulations with the parameters estimated from the empirical data \cite{davidsen17}: $\theta_1=0.25$, $\theta_2=0.7$ and a transition point $\tau_{c}=10^{\alpha_{\tau} m_{0}}x_{c}$. The magnitudes of the events are generated from Eq.~\eqref{eq:BD:gr} with $b=2.0$ and $m_{c}=3.25$ \footnote{Please note that the magnitudes of AE events in the lab experiments are defined in a different way than for earthquakes and not directly comparable. Hence the difference in scales and $b$-values.}. The productivity exponent of the kernel is set to $\alpha=0.55$ and $\alpha_{\tau}=0.5$. Thus the average number of events generated by a mainshock is $\langle N_{AS}(m_{0})\rangle = k 10^{1.05 m_{0}}$. Combining Eqs.~\eqref{eq:BD:scaling} and \eqref{eq:BD:stdETAS}, the explicit dependence of the productivity term on the branching ratio $n_{b}$ reads: 

\begin{equation}
    k(n_{b})=\int dm_{0}  \int{d\tau}\phi(\tau| m_{0}) = n_{b}\left({1-\frac{\alpha-\alpha_{\tau}}
    {b}}\right)10^{-(\alpha+\alpha_{\tau})m_{c}}
\end{equation}. 

We generated sequences of $10^{5}$ background events for different values of $x_{c}$ and $n_{b}$. To highlight our main findings, we focus on two specific examples in the following: Low branching ratio with $n_{b}=0.2$ and high branching ration with $n_{b}=0.95$. The ratio between the transition point and average activity of unrelated events determines whether the second power-law regime can be observed in the mean aftershock sequence rates or not.
For example, the transition point $x_{c}$ is unobservable when the triggering rates $\phi(x_{c})$ fall below the background rate $\mu_{0}$. 
Considering the parameters of our simulations, this happens for  $x_{c}=\frac{\theta_{1}\theta_{2}}{\theta_{1}+\theta_{2}}= 5/300 \sim 0.0167$ in our reduced units.
We simulate the model with values of $x_{c}=1, 5/300, 10^{-8}$ such that the transition point is found above, around and below the background level.
In order to evaluate the effect of time-independent vs time-dependent background rates, we impose $\mu_{0}(t) \sim t^{\sigma-1}$ by sampling the background events from a cumulative distribution: $\mathrm{CDF}(t_{i})=t^{\sigma}$. Thus, a constant rate is sampled for $\sigma=1$ and quadratic increasing rate for $\sigma=3$, resembling the smooth increase of the rate observed at the beginning of AE experiments~\cite{davidsen17}.\\

Fig.~\ref{fig:BD:M1} shows the measured bare rates for all simulations to verify the scaling relation~\eqref{eq:BD:scaling}. Each line represents an average over all parent events with magnitude $m < M_i < m + \Delta m$ in each simulation. Indeed, the numerical results reproduce the expected relation from Eq.~\eqref{eq:BD:Kernel}, represented by the wide grey curve. \\

\begin{figure}
\centering 
\includegraphics{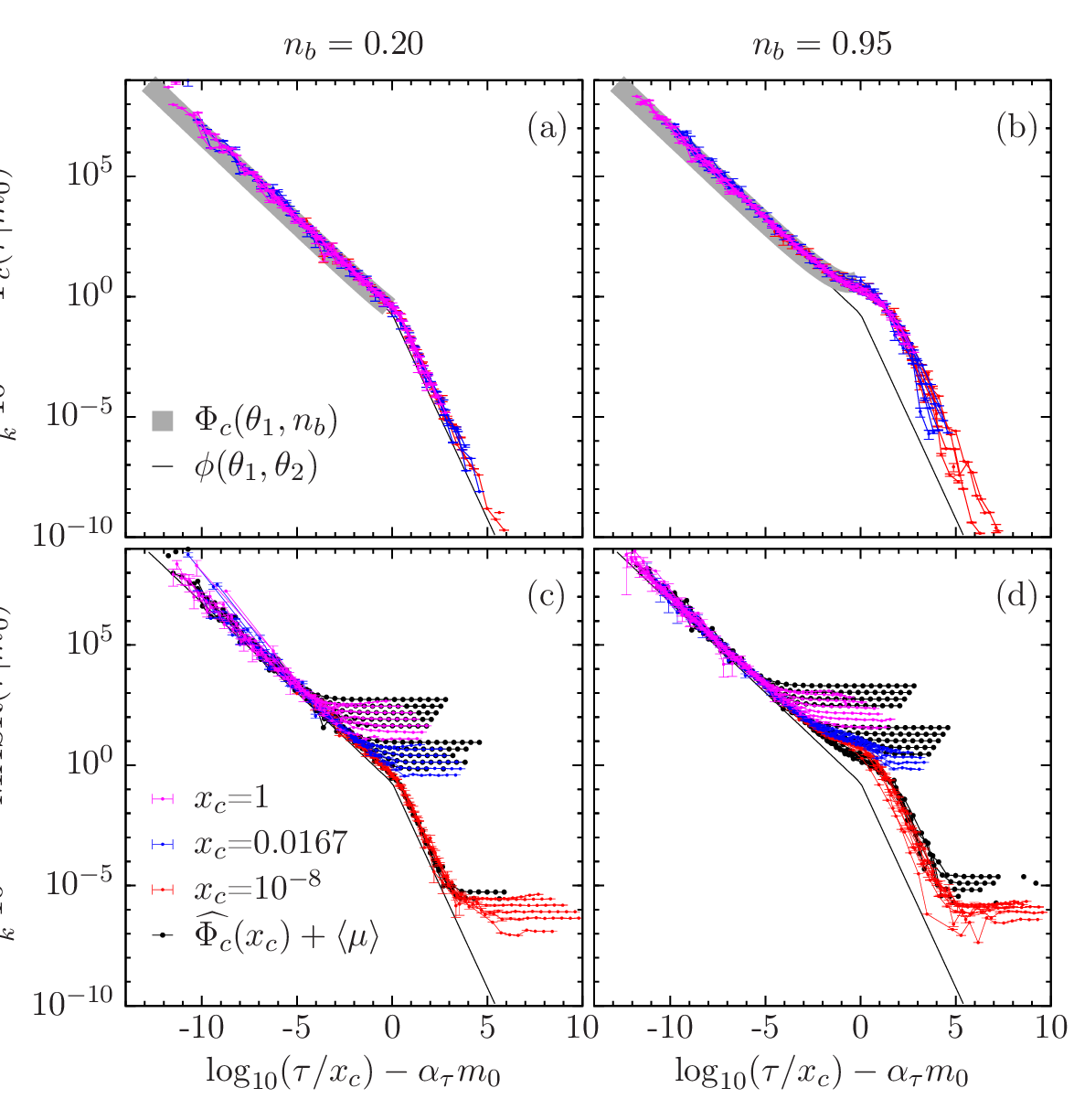}
\caption{Rescaled \emph{compound} triggering rates (a,b) and mean aftershock sequence rates (c,d) for the same simulations and mainshock magnitude bins as in Fig.~\ref{fig:BD:M1}. The scaling function for the triggering kernel is shown as thin black lines and the numerical solution of Eq.~\eqref{eq:BD:M2theta1} is shown in grey. In (c,d) the measured compound rates plus the average rate $\langle \mu \rangle$ from Eq.~\eqref{eq:BD:M3} are also plotted (black dots and lines). Error bars in (c,d) correspond to one standard deviation for each individual sequence.
The rate of events unrelated to the given triggering cascade $\langle \mu(t) \rangle$ (see Eq.~\eqref{eq:BD:M3}) decreases with the mainshock magnitude in the rescaled representation due to the term $10^{-\alpha m_{0}}$, giving rise to the variations in the plateaus observed at late times. }
\label{fig:BD:M23}       
\end{figure}

The compound triggering rates are also invariant with respect to $\tau_{c}=10^{-\alpha_{\tau} m_{0}}x_{c}$, which controls the temporal scale of the triggering with respect to the background rate, but it is sensitive to the productivity (given by $\int \phi(\tau| m_{0}) dm_{0}$), as stated in Eq.~\eqref{eq:BD:GN}, and, thus, depends implicitly on $n_{b}$. The top panels of Fig.~\ref{fig:BD:M23} show the average of the compound rates in mainshock magnitude windows of $\Delta m = 0.5$, measured in the numerical simulations. The curves are scaled according to Eq.~\eqref{eq:BD:scaling} in order to identify the deviations from the direct rates. For low branching ratio ($n_{b}=0.2$) the  compound rates are almost indistinguishable from the direct rates. This is not the case for higher branching ratios ($n_{b}=0.95$) where we identify an exceedance of activity starting at $\tau\sim \tau_{c}$ and extending to higher values. 
Eq.~\eqref{eq:BD:M2solution} is only valid if the dependence on mainshock magnitude can be separated from the temporal kernel. In the modified ETAS model, the coupling imposed in the temporal scale $\tau_c(m_{0})$ prevents this analytical approach. Yet, if we limit our analysis to the short-time power-law regime only (below $\tau_c$), we can at least provide an explanation for the exceedance point observed for $n_b=0.95$ in Fig.~\ref{fig:BD:M23}. 
If the triggering kernel consisted of a single power-law regime with exponent $\theta_{1}>0$ at all time-scales $\tau \to \infty$, the compound rates would always diverge and increase exponentially fast above a certain characteristic time $\tau^{*}$.
When the branching ratio is high, the characteristic time $\tau^{*}$ is reached before the transition time $\tau_{c}$ towards the fast decaying regime.
Since we now only consider the term: $\phi(\tau/\tau_{c}) = \left({\frac{\tau}{\tau_{c}}}\right)^{\theta_{1}-1}$, we can separate the productivity from the pure temporal kernel and, following Eq.~\eqref{eq:BD:M2solution}, obtain the resulting rate for $\tau \leq \tau_{c}$:
\begin{equation}
        \Phi_{c} ({\tau/\tau_{c}})= 
   \frac{k}{x_{c}} 10^{\alpha m_{0}}\frac{\theta_{1}\theta_{2}}{\theta_{2}+\theta_{1}}\Gamma(\theta_{1}) 
   \sum_{g=1}^{\infty}\frac{1}{\Gamma (g \theta_{1})}
   \left({
   n_{b}\left({\frac{\tau}{\tau_{c}}}\right)^{\theta_{1}}
   }\right)^{g} 
   \label{eq:BD:M2theta1}
\end{equation}
with $\Gamma(\beta):= \int_{0}^{\infty} t^{\beta-1}\exp(-t)dt$. In Fig.~\ref{fig:BD:M23}.a,b the numerical solution from Eq.~\eqref{eq:BD:M2theta1} (computed up to $g=50$) is plotted as a guide to the eye, revealing the trend to the deviation from Eq.~\eqref{eq:BD:scaling} for high branching ratios. The solution for any value of $\theta_{2}>0$ and $n_{b}<1$ above the transition point $\tau=\tau_{c}$  should not differ significantly from the power-law decay discussed in Ref.~\cite{Helmstetter2002}.\\

\begin{figure}
\centering 
\includegraphics{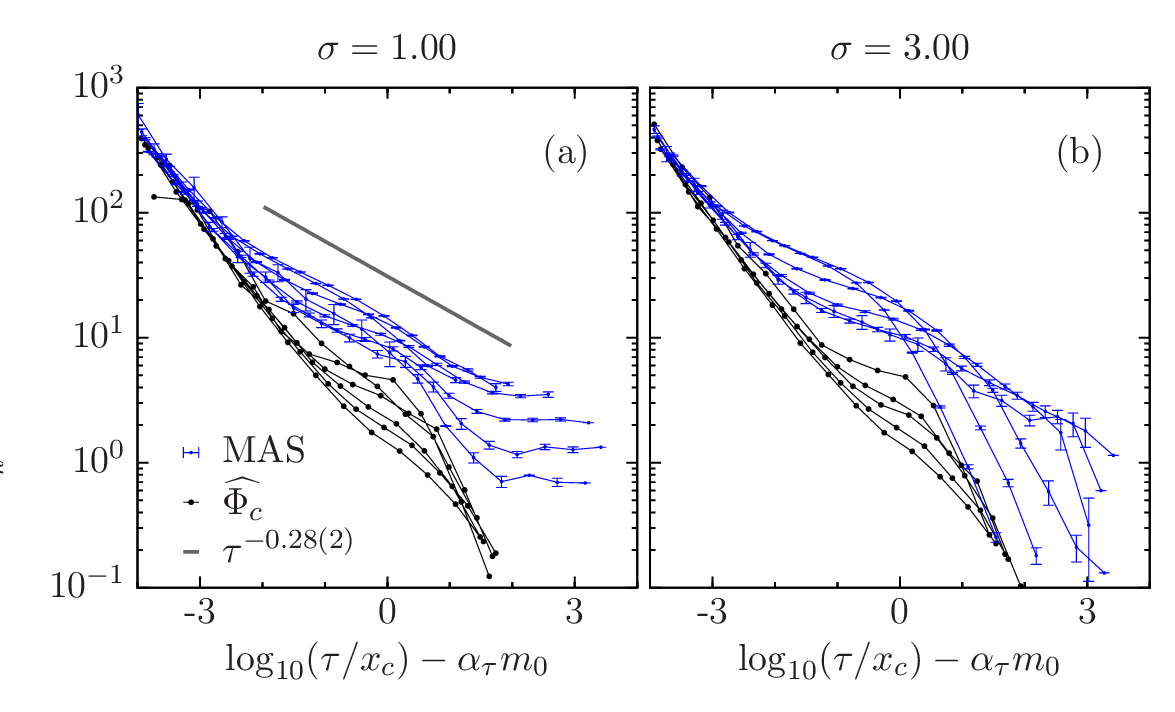}
\caption{(a) Rescaled MASR (blue) and compound rates ($\widehat{\Phi_{c}}$ in black) in a short interval around $\tau \sim \tau_{c}$ for $x_{c}=0.0167$ and $n_{b}=0.95$. The thick line represents a power-law with an average estimated value of $p=0.28(2)$ for the mean aftershock sequence rates. (b) Same as in (a) but for a time evolving background rate with a parabolic increase ($\sigma=3$). }
\label{fig:BD:M3}       
\end{figure}

\section{Inferring triggering rates in the modified ETAS model}

Without spatial information, the catalogs generated from a point process are given in a sequence of events $t_{i},m_{i}$ and the direct and compound triggering rates cannot be measured directly. Instead, we have to rely on the measurement of mean aftershock sequence rates (MASR) to infer the triggering kernel. In this section we present the results for synthetic catalogs generated from the modified ETAS model (Eq.~\eqref{eq:BD:Kernel}), and compare them with the actual direct and compound triggering rates.\\

Due to the increase in the aftershock rates with the magnitude of the mainshock as expressed by the productivity law, the overall intensity of the process as defined in Eq.~\eqref{eq:BD:hawkes} is typically dominated by an earlier large event or a more recent smaller one. To take advantage of this, we evaluate the mean aftershock sequence rates (MASR) as the activity after any event of any magnitude $M_{i}$ until the next event $j$ with magnitude $M_{j}>M_{i}$, in hopes to obtain long triggering sequences with reliable information. To obtain sufficient statistics, we average the MASR in mainshock magnitude windows ($m < M_i < m + \Delta m$). We must normalize the sequences by the measurement range of each sequence: the time until an event is found with $M_{j}>M_{i}$, equivalent to the complementary cumulative distribution of waiting times $\mathrm{CCDF}(\delta, M_{k}>m)$. In Fig.~\ref{fig:BD:M23}.c,d the MASR measured in simulations are compared to the compound triggering rates ($\Phi_{c} (\tau, m)$) with the addition of the \emph{time-averaged} rate $\overline{\langle \mu (t) \rangle}$, presenting an approximation to the final expression of Eq.~\eqref{eq:BD:M3}. Although the approximation is especially well-suited for low branching ratios, at $n_{b}\sim 1$, this assumption overestimates the contribution of unrelated events $\langle \mu(t) \rangle$ at late times. This is due to taking the time average since 
$\langle \mu (t) \rangle$ can significantly vary over time. Since triggered events are considered as mainshocks in the MASR, the rate of events triggered by independent branches within the same triggering tree are non-negligible. Although not directly triggered, this type of activity occurs predominantly at early times after the mainshock leading to a time-varying $\langle \mu (t) \rangle$. In general, we are able to recognize the double power-law kernel as long as the average rate $\langle \mu (t) \rangle$ is much lower than the triggering rate at the transition point $\frac{\theta_{1}+\theta_{2}}{\theta_{1}\theta_{2}}x_{c}$. Around and above this value, the average rate makes the secondary power-law regime unobservable and renders a single power-law decay in triggering rates resembling the standard Omori Utsu relation, but with an exponent value that is unphysically low, as was found in Refs.~\cite{baro2013,baro2014,ribeiro15}.\\

In the case of $\langle \mu (t) \rangle \sim \frac{\theta_{1}+\theta_{2}}{\theta_{1}\theta_{2}}x_{c}$ and $n_{b}\sim 1$,  MASR (blue lines in Fig.~\ref{fig:BD:M23}.d) render an effective power-law behavior with a low exponent extending up to 5 decades around $x_{c}$. This behavior is a consequence of the interplay between the transition point,  the rate of independent events, and the contribution of higher generation triggering. Fig.~\ref{fig:BD:M3} shows this region in more detail for both a simulation with constant background rate (Fig.~\ref{fig:BD:M3}.a) and for $\sigma=3$ (Fig.~\ref{fig:BD:M3}.b). In the MASR for $\sigma=1$, we can fit the effective power-law with an exponent lower than 0.3, not directly related to $\theta_{1}$ nor $\theta_{2}$, nor observed in the compound rates. The scaling relations of the triggering rates with $m_{0}$ are also affected. Thus, a blind estimation of Omori ($p$) and the productivity ($\alpha$) exponents limited within this interval by fitting and collapsing the MASR curves according to the scaling relations is unlikely to retrieve the right form of the triggering kernel.\\

\begin{figure}
\centering
\includegraphics{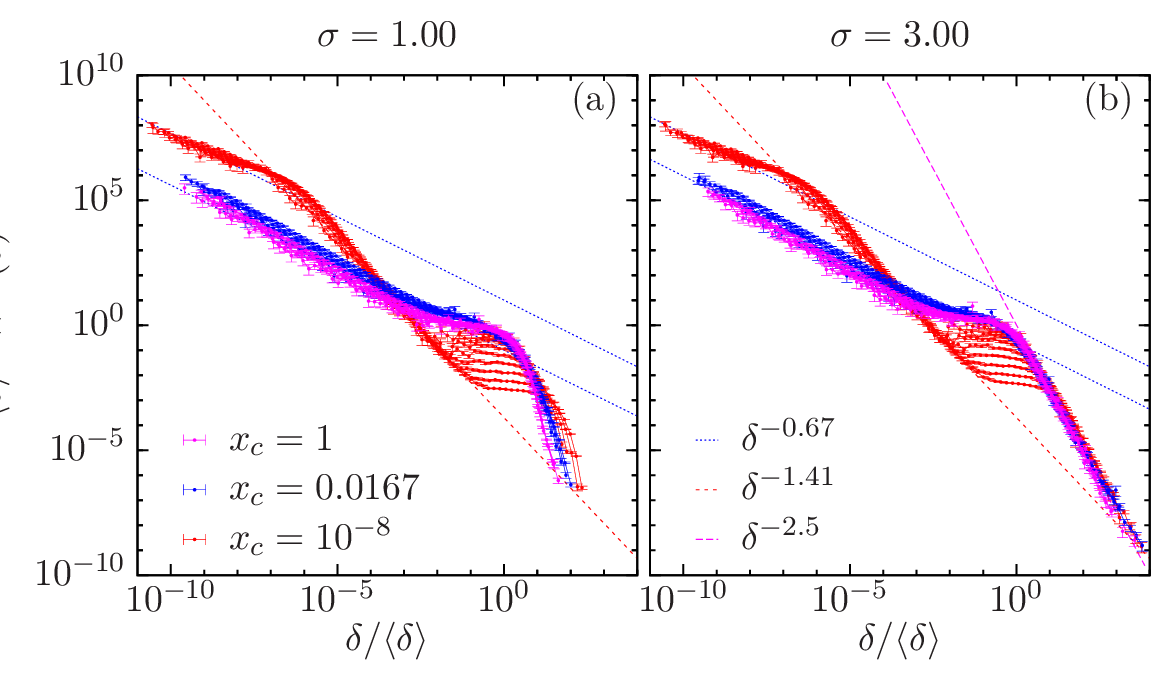}
\caption{Distribution of waiting times rescaled by the average value for $n_{b}=0.95$ and $x_{c}$ above, below and at the background level, and for (a) a uniform background rate ($\sigma=1$) and (b) for a background rate parabolically rising from $\mu_{0} = 0$ ($\sigma=3$). Each line represents a different magnitude threshold $m_{c}=3.25,3.50, ..., 6.5$. The expected power-law behaviors are represented as a guide to the eye. One sigma error bars are shown.}
\label{fig:BD:WT}       
\end{figure}

Finally, we evaluate the distribution of waiting times (DWT) for different magnitude thresholds, as shown in Fig.~\ref{fig:BD:WT}. We compare the results obtained with (a) the uniform ($\sigma=1$) and (b) time-dependent ($\sigma=3$) background rate, for different values of $x_{c}$ and $n_{b}=0.95$. No significant differences are found for $n_{b}=0.20$ (not shown). The distributions are scaled with the mean waiting time for each threshold, following the scaling relation expected for a Poisson process and observed also for other processes such as seismicity~\cite{corral04,davidsen2013} and rock fracture~\cite{davidsen05sg}, for example. In the standard ETAS model, the situation is more complicated~\cite{touati11,lippiello12}. This is also true for the modified ETAS model we consider here. Due to the additional scaling parameter ($\alpha_{\tau}$) the collapsing of the curves to a single scaling function can only be fulfilled over certain ranges. If the events were independent, from the background sampling one would expect an exponential distribution in the DWT for $\sigma=1$ and a power law decay with exponent: $\frac{2\sigma -1}{\sigma -1}=2.5$ for $\sigma=3$. Instead, both behaviors are only found for waiting times longer than the typical waiting time of background events  ($\delta \sim \langle \delta \rangle $). Below these times, the distribution is dominated by the triggering process, returning the power-law exponents predicted from the relation $2-1/p$: $0.67$ for the regime with $p=0.75$ and $1.41$ for the regime $p=1.7$, see Fig.~\ref{fig:BD:WT}. \\


\section{Discussion}

As expected, the branching ratio, the ratio between the power-law transition and the background rate are essential to understand the results of MASR in terms of the triggering kernel. 
The existence of a characteristic scale in the temporal triggering kernel offers a plausible explanation to the detection of effective Omori exponents lower than one in unlocalized catalogs of acoustic emission during mechanical processes \cite{baro2013,nataf2014,baro2016,ribeiro15,makinen2015,costa2016} and calorimetry  in structural phase transitions \cite{baro2014}. 
In the specific case of the failure of porous materials under compression~\cite{baro2013,nataf2014} an effective Omori exponent $p\sim 0.7$ was observed using MASR, compatible with the short time power-law regime found in localized catalogs \cite{davidsen17} and the MASR of the modified ETAS model (Fig.~\ref{fig:BD:M23}) for transition values $x_{c} \gtrsim \frac{\theta_{1}\theta_{2}}{\theta_{1}+\theta_{2}}$. The explanation derived from the modified ETAS models is that the second power law regime is hidden by the background rate. As a consequence, the estimation of the productivity term $k(m_{0})$ will neglect the existance of a scaling relation in the temporal axis. Thus, only a single exponent can be estimated by collapsing the curves $\phi(\tau|m_{0})$. In the modified ETAS model, if the second power-law is not observed in the MASR, one would find a scaling relation $k(m_{0}) \sim 10^{\alpha' m_{0}}$, but this measured exponent $\alpha'$ does not correspond to the scaling parameter $\alpha$, nor the productivity exponent $\widetilde{\alpha}=\alpha+\alpha_{\tau}$. Instead it corresponds to an intermediate value $\alpha'=\alpha+\alpha_{\tau}(1-\theta_{1})= \widetilde{\alpha} - \theta_{1}\alpha_{\tau}$. This relation directly follows from Eqs.~\eqref{eq:BD:Kernel} and~\eqref{eq:BD:scaling} if one only considers the regime $x\leq x_{c}$. The specific value of $\alpha'=0.925$ for our simulations (and consistent with the experiments in~\cite{davidsen17}) is, however, different from the value $\alpha'\approx 0.5$ observed during the failure of porous materials under compression~\cite{baro2013} using MASR. Provided the validity of the modified ETAS model, this suggests that the exponents $\alpha$ and $\alpha_{\tau}$ are not universal across rock fracture experiments. Assuming positive values of $\alpha$, $\alpha_{\tau}$ and $\theta_1=0.25$, both exponents are however limited within the range $0\leq \alpha \leq 2$ and $0\leq \alpha_{\tau} \leq 2/3$. \\

The distribution of waiting times in experimental data usually exhibits a sharp transition between two power-law regimes \cite{corral2003,baro2013,makinen2015}. 
The power-law regime observed for long waiting times is consistent with the temporal variations of the background rate. Our modified ETAS model can reproduce both observations if the maximum background rate is comparable to the rate at the transition point between the two power laws in the triggering kernel, i.e. $x_c = 0.0167$ in our simulations.  The long time regime of the triggering kernel is only observable before the regime dominated by the background rate for $x_c < 0.0167$, as shown in Fig.~\ref{fig:BD:WT}. The absence of the secondary  triggering  regime for $x_c \gtrsim 0.0167$  is also consistent with the above mentioned absence of the secondary regime above the background rate in the experimentally measured MASR.  \\

Finally, the exact mathematical relation between the $p$-value of the direct triggering rates and the power-law exponent in the waiting time distribution ($2-1/p$) is not consistent with the exponents measured using MASR in some experiments~\cite{baro2013,makinen2015}. While one might expect that this is related to the presence of a high branching ratio~\cite{Helmstetter2002}, this is not supported by our findings for the modified ETAS model considered here and remains an open question.

\section{Conclusions}
The measurement of triggering rates is a non-trivial problem, even in simplified branching processes. Due to limitations in the acquisition systems, we often must rely on the indirect measurement of mean aftershock sequence rates (MASR) to infer the original triggering kernel. The performance of this technique will depend specifically on the background rate and the branching ratio. We can retrieve the triggering kernel with a good precision whenever the individual triggering trees can be separated at low background rates and if secondary triggering can be neglected due to low branching ratio. But, in general, one should expect a strong superposition of independent and secondary activity. \\

When the data is sampled from a triggering kernel with characteristic time scales --- such as proposed in~\cite{davidsen17} for rock fracture and implemented in this work in the form of a modified ETAS model --- the interplay between the characteristic scale of the triggering and the background rate can render non-scaling regimes in both the measurement of triggering rates using MASR and the distribution of waiting times. 
As a specific case, if the characteristic time scale and the time scale of the background activity are comparable we find a crossover regime similar to another power law. Yet, its exponent cannot be trivially associated with the underlying parameters of the modified ETAS model. \\

In more general terms, our study here shares light on the problem of separating overlapping triggering cascades or branching trees from limited information to establish the underlying causal relationships, which is not specific to slip and fracture events. Indeed, another prime example is neuronal activity for which such investigations are still at the very beginning~\cite{williams2016,orlandi2013}.
 
\bibliography{allbib,j3}{}

\begin{thebibliography}{10}

\bibitem{sethna01}
J.~P. Sethna, K.~A. Dahmen, and C.~R. Myers.
\newblock Crackling noise.
\newblock {\em Nature (London)}, 410:242, 2001.

\bibitem{durin06}
G.~Durin and S.~Zapperi.
\newblock In G.~Bertotti and I.~Mayergoyz, editors, {\em The Science of
  Hysteresis}, volume II., page 181. Academic Press, San Diego, 2006.

\bibitem{papanikolaou11}
S.~Papanikolaou, F.~Bohn, R.~L. Sommer, G.~Durin, S.~Zapperi, and J.~P. Sethna.
\newblock Universality beyond power laws and the average avalanche shape.
\newblock {\em Nature Physics}, 7:316, 2011.

\bibitem{baro2014}
Jordi Bar{\'{o}}, Jos{\'{e}}-Mar{\'{i}}a Mart{\'{i}}n-Olalla, Francisco~Javier
  Romero, Mar{\'{i}}a~Carmen Gallardo, Ekhard K~H Salje, Eduard Vives, and
  Antoni Planes.
\newblock {Avalanche correlations in the martensitic transition of a Cu--Zn--Al
  shape memory alloy: analysis of acoustic emission and calorimetry}.
\newblock {\em Journal of Physics: Condensed Matter}, 26(12):125401, 2014.

\bibitem{bonamy11}
D.~Bonamy and E.~Bouchaud.
\newblock Failure of heterogeneous materials: A dynamic phase transition?
\newblock {\em Physics Reports}, 498:1, 2011.

\bibitem{tantot13}
A.~Tantot, S.~Santucci, O.~Ramos, S.~Deschanel, M.-A. Verdier, E.~Mony, Y.~Wei,
  S.~Ciliberto, L.~Vanel, and P.~C.~F. Di~Stefano.
\newblock Sound and light from fractures in scintillators.
\newblock {\em Physical Review Letters}, 111:154301, 2013.

\bibitem{baro2013}
Jordi Bar{\'{o}}, Steve Dixon, Rachel~S Edwards, Yichao Fan, Dean~S Keeble,
  Llu{\'{i}}s Ma{\~{n}}osa, Antoni Planes, and Eduard Vives.
\newblock {Simultaneous detection of acoustic emission and Barkhausen noise
  during the martensitic transition of a Ni-Mn-Ga magnetic shape-memory alloy}.
\newblock {\em Physical Review B}, 88(17):174108, 2013.

\bibitem{kun14}
F.~Kun, I.~Varga, S.~Lennartz-Sassinek, and I.~G. Main.
\newblock Rupture cascades in a discrete element model of a porous sedimentary
  rock.
\newblock {\em Physical Review Letters}, 112:065501, 2014.

\bibitem{maekinen15}
T.~M\"akinen, A.~Miksic, M.~Ovaska, and Mikko~J. Alava.
\newblock Avalanches in wood compression.
\newblock {\em Physical Review Letters}, 115:055501, 2015.

\bibitem{ribeiro15}
H.~V. Ribeiro, L.~S. Costa, L.~G.~A. Alves, P.~A. Santoro, S.~Picoli, E.~K.
  Lenzi, and R.~S. Mendees.
\newblock Analogies between the cracking noise of ethanol-dampened charcoal and
  earthquakes.
\newblock {\em Physical Review Letters}, 115:025503, 2015.

\bibitem{benzion08}
Y.~Ben-Zion.
\newblock Collective behavior of earthquakes and faults: Continuum-discrete
  transitions, progressive evolutionary changes, and different dynamic regimes.
\newblock {\em Review of Geophysics}, 46:RG4006, 2008.

\bibitem{gu13}
C.~Gu, G.~St-Yves, and J.~Davidsen.
\newblock Spiral wave chimeras in complex-oscillatory and chaotic systems.
\newblock {\em Physical Review Letters}, 111:134101, 2013.

\bibitem{davidsen16m}
J{\"o}rn Davidsen and Marco Baiesi.
\newblock Self-similar aftershock rates.
\newblock {\em Physical Review E}, 94(2):022314, 2016.

\bibitem{maghsoudi16}
Samira Maghsoudi, David~W Eaton, and J{\"o}rn Davidsen.
\newblock Nontrivial clustering of microseismicity induced by hydraulic
  fracturing.
\newblock {\em Geophysical Research Letters}, 43(20), 2016.

\bibitem{baiesi06m}
M.~Baiesi, M.~Paczuski, and A.~L. Stella.
\newblock Intensity thresholds and the statistics of the temporal occurrence of
  solar flares.
\newblock {\em Physical Review Letters}, 96:051103, 2006.

\bibitem{arcangelis06a}
L.~de~Arcangelis, C.~Godano, E.~Lippiello, and M.~Nicodemi.
\newblock Universality in solar flare and earthquake occurrence.
\newblock {\em Physical Review Letters}, 96:051102, 2006.

\bibitem{moloney11}
N.~R. Moloney and J.~Davidsen.
\newblock Extreme bursts in the solar wind.
\newblock {\em Geophysical Research Letters}, 38:L14111, 2011.

\bibitem{moloney13}
N.~R. Moloney and J.~Davidsen.
\newblock Stationarity of extreme bursts in the solar wind.
\newblock {\em Physical Review E}, page submitted, 2013.

\bibitem{turcotte99}
D.~L. Turcotte.
\newblock Self-organized criticality.
\newblock {\em Reports on Progress in Physics}, 62:1377, 1999.

\bibitem{bak}
P.~Bak.
\newblock {\em How nature works}.
\newblock {Copernicus}, New York, 1996.

\bibitem{crutchfield}
J.~P. Crutchfield and P.~Schuster.
\newblock {\em Evolutionary Dynamics: Exploring the Interplay of Selection,
  Accident, Neutrality, and Function}.
\newblock Oxford University Press US, NY, NY, 2003.

\bibitem{drossel01}
B.~Drossel.
\newblock Biological evolution and statistical physics.
\newblock {\em Advances in Physics}, 50:209, 2001.

\bibitem{friedman12}
N.~Friedman, S.~Ito, B.~A.~W. Brinkman, M.~Shimono, R.~E.~L. DeVille, K.~A.
  Dahmen, J.~M. Beggs, and T.~C. Butler.
\newblock Universal critical dynamics in high resolution neuronal avalanche
  data.
\newblock {\em Physical Review Letters}, 108:208102, 2012.

\bibitem{yaghoubi17}
Mohammad Yaghoubi, Ty~de~Graaf, Javier~G. Orlandi, Fernando Girotto, Michael
  Colicos, and J\"orn Davidsen.
\newblock Non-trivial neuronal avalanche dynamics in developing neuronal
  cultures.
\newblock {\em (preprint)}, 2017.

\bibitem{sornette1996}
Didier Sornette, Anders Johansen, and Jean-Philippe Bouchaud.
\newblock Stock market crashes, precursors and replicas.
\newblock {\em Journal de Physique I}, 6(1):167--175, 1996.

\bibitem{farmer05}
J.~D. Farmer, D.~E. Smith, and M.~Shubik.
\newblock Is economics the next the next physical sciences?
\newblock {\em Physics Today}, 58(9):37, 2005.

\bibitem{lillo03}
F.~Lillo and R.~N. Mantegna.
\newblock Power-law relaxation in a complex system: Omori law after a financial
  market crash.
\newblock {\em Physical Review E}, 68:016119, 2003.

\bibitem{weber07}
P.~Weber, F.~Wang, I.~Vodenska-Chitkushev, S.~Havlin, and H.~E. Stanley.
\newblock Relation between volatility correlations in financial markets and
  {Omori} processes occurring on all scales.
\newblock {\em Physical Review E}, 76:016109, 2007.

\bibitem{petersen10}
A.~M. Petersen, F.~Wang, S.~Havlin, and H.~E. Stanley.
\newblock Market dynamics immediately before and after financial shocks:
  Quantifying the {O}mori, productivity, and {B}ath laws.
\newblock {\em Physical Review E}, 82:036114, 2010.

\bibitem{siokis12}
F.~M. Siokis.
\newblock Stock market dynamics: Before and after stock market crashes.
\newblock {\em Physica A}, 391:1315, 2012.

\bibitem{klimek11}
P.~Klimek, W.~Bayer, and S.~Thurner.
\newblock The blogosphere as an excitable social medium: Richter’s and
  omori’s law in media coverage.
\newblock {\em Physica A}, 390:3870, 2011.

\bibitem{weiss04}
J.~Weiss and M.~Carmen Miguel.
\newblock Dislocation avalanche correlations.
\newblock {\em Materials Science and Engineering A}, 387--389:292, 2004.

\bibitem{crane08}
R.~Crane and D.~Sornette.
\newblock Robust dynamic classes revealed by measuring the response function of
  a social system.
\newblock {\em Proceedings National Academy of Sciences U.S.A.}, 105:15649,
  2008.

\bibitem{sornette09}
D.~Sornette and S.~Utkin.
\newblock Limits of declustering methods for disentangling exogenous from
  endogenous events in time series with foreshocks, main shocks, and
  aftershocks.
\newblock {\em Physical Review E}, 79:061110, 2009.

\bibitem{gu13amj}
C.~Gu, A.~Y. Schumann, M.~Baiesi, and J.~Davidsen.
\newblock Triggering cascades and statistical properties of aftershocks.
\newblock {\em Journal of Geophysical Research}, 118:4278, 2013.

\bibitem{hainzl14}
S.~Hainzl, J.~Moradpour, and J.~Davidsen.
\newblock Static stress triggering explains the empirical aftershock distance
  decay.
\newblock {\em Geophysical Research Letters}, 41:8818, 2014.

\bibitem{stojanova14}
M.~Stojanova, S.~Santucci, L.~Vanel, and O.~Ramos.
\newblock High frequency monitoring reveals aftershocks in subcritical crack
  growth.
\newblock {\em Physical Review Letters}, 112:115502, 2014.

\bibitem{utsu95}
T.~Utsu, Y.~Ogata, and R.~S. Matsu'ura.
\newblock The centenary of the {Omori} formula for a decay law of aftershock
  activity.
\newblock {\em J. Phys. Earth}, 43:1, 1995.

\bibitem{moradpour14}
J.~Moradpour, S.~Hainzl, and J.~Davidsen.
\newblock Nontrivial decay of aftershock density with distance in {Southern
  California}.
\newblock {\em Journal of Geophysical Research}, 119:5518, 2014.

\bibitem{davidsen14}
J.~Davidsen, C.~Gu, and M.~Baiesi.
\newblock Generalized {Omori-Utsu} law for aftershock sequences in southern
  {California}.
\newblock {\em Geophysical Journal International}, 201:965, 2015.

\bibitem{goebel13a}
T.~H.~W. Goebel, C.~G. Sammis, T.~W. Becker, G.~Dresen, and D.~Schorlemmer.
\newblock A comparision of seismicity characteristics and fault structure in
  stick-slip experiments and nature.
\newblock {\em Pure and Applied Geophysics}, 172:2247, 2013.

\bibitem{omori}
F.~Omori.
\newblock On the aftershocks of earthquakes.
\newblock {\em Journal of College Science, Imperial University of Tokyo},
  7:111, 1894.

\bibitem{castillo2013}
Pedro~O Castillo-Villa, Jordi Bar{\'{o}}, Antoni Planes, Ekhard K~H Salje,
  Pathikumar Sellappan, Waltraud~M Kriven, and Eduard Vives.
\newblock {Crackling noise during failure of alumina under compression: the
  effect of porosity}.
\newblock {\em Journal of Physics: Condensed Matter}, 25(29):292202, 2013.

\bibitem{nataf2014}
Guillaume~F. Nataf, Pedro~O. Castillo-Villa, Jordi Bar{\'{o}}, Xavier Illa,
  Eduard Vives, Antoni Planes, and Ekhard K.~H. Salje.
\newblock {Avalanches in compressed porous Si02-based materials}.
\newblock {\em Physical Review E}, 90(2):022405, 2014.

\bibitem{davidsen05sg}
J.~Davidsen, S.~Stanchits, and G.~Dresen.
\newblock Scaling and universality in rock fracture.
\newblock {\em Physical Review Letters}, 98:125502, 2007.

\bibitem{davidsen2013}
J{\"{o}}rn Davidsen and Grzegorz Kwiatek.
\newblock {Earthquake Interevent Time Distribution for Induced Micro-, Nano-,
  and Picoseismicity}.
\newblock {\em Physical Review Letters}, 110(6):68501, feb 2013.

\bibitem{goodfellow2014}
SD~Goodfellow and RP~Young.
\newblock A laboratory acoustic emission experiment under in situ conditions.
\newblock {\em Geophysical Research Letters}, 41(10):3422--3430, 2014.

\bibitem{davidsen17}
Elli-Maria Charalampidou Thomas Goebel Sergei Stanchits Marc~R\"uck
  J\"orn~Davidsen, Grzegorz~Kwiatek and Georg Dresen.
\newblock Triggering proccesses in rock fracture.

\bibitem{davidsen06pm}
J.~Davidsen, P.~Grassberger, and M.~Paczuski.
\newblock Networks of recurrent events, a theory of records, and application to
  finding causal signatures in seismicity.
\newblock {\em Physical Review E}, 77:066104, 2008.

\bibitem{marsan08}
D.~Marsan and O.~Lenglin\'e.
\newblock Extending earthquake' reach through cascading.
\newblock {\em Science}, 319:1076, 2008.

\bibitem{zaliapin13a}
I.~Zaliapin and Y.~Ben-Zion.
\newblock Earthquake clusters in southern {California}, {I}: Identification and
  stability.
\newblock {\em Journal of Geophysical Research}, 118:2847, 2013.

\bibitem{helmstetter2003}
Agn{\`{e}}s Helmstetter.
\newblock {Is Earthquake Triggering Driven by Small Earthquakes?}
\newblock {\em Physical Review Letters}, 91(5):58501, 2003.

\bibitem{kagan87}
Y.~Y. Kagan and L.~Knopoff.
\newblock Statistical short-term earthquake prediction.
\newblock {\em Science}, 236:1563, 1987.

\bibitem{ogata1999}
Yosihiko Ogata.
\newblock Seismicity analysis through point-process modeling: A review.
\newblock In {\em Seismicity patterns, their statistical significance and
  physical meaning}, pages 471--507. Springer, 1999.

\bibitem{daley2007}
Daryl~J Daley and David Vere-Jones.
\newblock {\em {An introduction to the theory of point processes: volume II:
  general theory and structure}}, volume~2.
\newblock Springer Science {\&} Business Media, 2007.

\bibitem{corral2009}
{\'{A}}lvaro Corral.
\newblock {Point-occurrence self-similarity in crackling-noise systems and in
  other complex systems}.
\newblock {\em Journal of Statistical Mechanics: Theory and Experiment},
  2009(01):P01022, 2009.

\bibitem{hawkes1974}
Alan~G Hawkes and David Oakes.
\newblock {A cluster process representation of a self-exciting process}.
\newblock {\em Journal of Applied Probability}, pages 493--503, 1974.

\bibitem{helmstetter2006}
Agnes Helmstetter, Yan~Y Kagan, and David~D Jackson.
\newblock Comparison of short-term and time-independent earthquake forecast
  models for southern california.
\newblock {\em Bulletin of the Seismological Society of America},
  96(1):90--106, 2006.

\bibitem{filimonov2012}
Vladimir Filimonov and Didier Sornette.
\newblock Quantifying reflexivity in financial markets: Toward a prediction of
  flash crashes.
\newblock {\em Physical Review E}, 85(5):056108, 2012.

\bibitem{tiampo12}
K.~F. Tiampo and R.~Shcherbakov.
\newblock {Seismicity-based earthquake forecasting techniques: Ten years of
  progress}.
\newblock {\em Tectonophysics}, {522}:{89}, {2012}.

\bibitem{bacry2015}
Emmanuel Bacry, Iacopo Mastromatteo, and Jean-Fran{\c{c}}ois Muzy.
\newblock Hawkes processes in finance.
\newblock {\em Market Microstructure and Liquidity}, 1(01):1550005, 2015.

\bibitem{sibani2005}
Paolo Sibani and H~Jeldtoft Jensen.
\newblock Intermittency, aging and extremal fluctuations.
\newblock {\em EPL (Europhysics Letters)}, 69(4):563, 2005.

\bibitem{van2013}
Nicholas~J van~der Elst, Emily~E Brodsky, and Thorne Lay.
\newblock Remote triggering not evident near epicenters of impending great
  earthquakes.
\newblock {\em Bulletin of the Seismological Society of America},
  103(2B):1522--1540, 2013.

\bibitem{Jagla2014}
E.~A. Jagla.
\newblock {Aftershock production rate of driven viscoelastic interfaces}.
\newblock {\em Physical Review E - Statistical, Nonlinear, and Soft Matter
  Physics}, 90(4):1--8, 2014.

\bibitem{janicevic2016}
Sanja Jani{\'c}evi{\'c}, Lasse Laurson, Knut~J{\o}rgen M{\aa}l{\o}y,
  St{\'e}phane Santucci, and Mikko~J Alava.
\newblock Interevent correlations from avalanches hiding below the detection
  threshold.
\newblock {\em Physical Review Letters}, 117(23):230601, 2016.

\bibitem{zaliapin08}
I.~Zaliapin, A.~Gabrielov, V.~Keilis-Borok, and H.~Wong.
\newblock Clustering analysis of seismicity and aftershock identification.
\newblock {\em Physical Review Letters}, 101:018501, 2008.

\bibitem{hainzl2016}
S~Hainzl, A~Christophersen, D~Rhoades, and D~Harte.
\newblock Statistical estimation of the duration of aftershock sequences.
\newblock {\em Geophysical Journal International}, 205(2):1180--1189, 2016.

\bibitem{Helmstetter2002}
Agn{\`{e}}s Helmstetter and Didier Sornette.
\newblock {Subcritical and supercritical regimes in epidemic models of
  earthquake aftershocks}.
\newblock {\em Journal of Geophysical Research: Solid Earth}, 107(B10):ESE
  10--1----ESE 10--21, 2002.

\bibitem{Helmstetter2003b}
Agn{\`{e}}s Helmstetter, Didier Sornette, and Jean-Robert Grasso.
\newblock {Mainshocks are aftershocks of conditional foreshocks: How do
  foreshock statistical properties emerge from aftershock laws}.
\newblock {\em Journal of Geophysical Research: Solid Earth (1978--2012)},
  108(B1), 2003.

\bibitem{van2010}
Nicholas~J Van Der~Elst and Emily~E Brodsky.
\newblock Connecting near-field and far-field earthquake triggering to dynamic
  strain.
\newblock {\em Journal of Geophysical Research: Solid Earth}, 115(B7), 2010.

\bibitem{bi1989}
Hongguang Bi, Gerhard B{\"{o}}rner, and Yaoquan Chu.
\newblock {Correlations in the absorption lines of the quasar Q0420-388}.
\newblock {\em Astronomy and Astrophysics}, 218:19--23, 1989.

\bibitem{carbone2006}
V~Carbone, L~Sorriso-Valvo, A~Vecchio, F~Lepreti, P~Veltri, P~Harabaglia, and
  I~Guerra.
\newblock {Clustering of polarity reversals of the geomagnetic field}.
\newblock {\em Physical Review Letters}, 96(12):128501, 2006.

\bibitem{baro2016}
Jordi Bar{\'{o}}, Antoni Planes, Ekhard K~H Salje, and Eduard Vives.
\newblock {Fracking and labquakes}.
\newblock {\em Philosophical Magazine}, 6435(September):1--11, 2016.

\bibitem{baro2016b}
Jordi Bar{\'o}, Peter Shyu, Siyuan Pang, Iwona~M Jasiuk, Eduard Vives,
  Ekhard~KH Salje, and Antoni Planes.
\newblock Avalanche criticality during compression of porcine cortical bone of
  different ages.
\newblock {\em Physical Review E}, 93(5):053001, 2016.

\bibitem{wheatland2000}
M~S Wheatland.
\newblock {The Origin of the Solar Flare Waiting-Time Distribution}.
\newblock {\em The AstroPhysical Journal Letters}, 536(2):L109, 2000.

\bibitem{corral2003}
Alvaro Corral.
\newblock Local distributions and rate fluctuations in a unified scaling law
  for earthquakes.
\newblock {\em Physical Review E}, 68(3):035102, 2003.

\bibitem{shcherbakov2005}
Robert Shcherbakov, Gleb Yakovlev, Donald~L Turcotte, and John~B Rundle.
\newblock {Model for the distribution of aftershock interoccurrence times}.
\newblock {\em Physical Review Letters}, 95(21):218501, 2005.

\bibitem{makinen2015}
T.~M{\"{a}}kinen, a.~Miksic, M.~Ovaska, and Mikko~J. Alava.
\newblock {Avalanches in Wood Compression}.
\newblock {\em Physical Review Letters}, 115(5):055501, 2015.

\bibitem{gutenberg1944}
Beno Gutenberg and Charles~F Richter.
\newblock {Frequency of earthquakes in California}.
\newblock {\em Bulletin of the Seismological Society of America},
  34(4):185--188, 1944.

\bibitem{kwiatek10}
G.~Kwiatek, K.~Plenkers, M.~Nakatani, Y.~Yabe, G.~Dresen, and {JAGUARS-Group}.
\newblock Frequency-magnitude characteristics down to magnitude --4.4 for
  induced seismicity recorded at {M}poneng gold mine, {South Africa}.
\newblock {\em Bulletin of the Seismological Society of America}, 100:1165,
  2010.

\bibitem{davidsen11}
J.~Davidsen and A.~Green.
\newblock Are earthquake magnitudes clustered?
\newblock {\em Physical Review Letters}, 106:108502, 2011.

\bibitem{davidsen12}
J.~Davidsen, G.~Kwiatek, and G.~Dresen.
\newblock No evidence of magnitude clustering in an aftershock sequence of
  nano- and picoseismicity.
\newblock {\em Physical Review Letters}, 108:038501, 2012.

\bibitem{ogata1988}
Yosihiko Ogata.
\newblock {Statistical models for earthquake occurrences and residual analysis
  for point processes}.
\newblock {\em Journal of the American Statistical Association}, 83(401):9--27,
  1988.

\bibitem{corral04}
\'A. Corral.
\newblock Long-term clustering, scaling, and universality in the temporal
  occurrence of earthquakes.
\newblock {\em Physical Review Letters}, 92:108501, 2004.

\bibitem{touati11}
S.~Touati, M.~Naylor, I.~G. Main, and M.~Christie.
\newblock Masking of earthquake triggering behavior by a high background rate
  and implications for epidemic-type aftershock sequence inversions.
\newblock {\em Journal of Geophysical Research}, 116:B03304, 2011.

\bibitem{lippiello12}
Eugenio Lippiello, {\'A}lvaro Corral, Milena Bottiglieri, Cataldo Godano, and
  Lucilla de~Arcangelis.
\newblock Scaling behavior of the earthquake intertime distribution: Influence
  of large shocks and time scales in the omori law.
\newblock {\em Physical Review E}, 86(6):066119, 2012.

\bibitem{costa2016}
Leandro~S Costa, Ervin~K Lenzi, Renio~S Mendes, and Haroldo~V Ribeiro.
\newblock Extensive characterization of seismic laws in acoustic emissions of
  crumpled plastic sheets.
\newblock {\em EPL (Europhysics Letters)}, 114(5):59002, 2016.

\bibitem{williams2016}
Rashid~V Williams-Garcia, John~M Beggs, and Gerardo Ortiz.
\newblock Unveiling causal activity of complex networks.
\newblock {\em arXiv preprint arXiv:1603.05659}, 2016.

\bibitem{orlandi2013}
Javier~G. Orlandi, Jordi Soriano, Enrique Alvarez-Lacalle, Sara Teller, and
  Jaume Casademunt.
\newblock {Noise focusing and the emergence of coherent activity in neuronal
  cultures}.
\newblock {\em Nature Physics}, 9(9):582--590, 2013.

\end{thebibliography}
\bibliographystyle{unsrt}

\end{document}